\documentclass[runningheads]{llncs}
\usepackage[T1]{fontenc}

\usepackage[misc]{ifsym}
\usepackage{times}
\usepackage{soul}
\usepackage{url}
\usepackage{graphicx}
\usepackage{amsmath}
\usepackage{booktabs}
\usepackage{algorithm}
\usepackage{algorithmic}
\usepackage{graphicx}
\usepackage{amssymb}
\usepackage{xcolor}
\usepackage{colortbl}
\usepackage{threeparttable}
\usepackage{float}
\usepackage{subfig}
\usepackage{newfloat}
\usepackage{listings}
\usepackage[colorlinks,
            linkcolor=red,
            anchorcolor=green,
            citecolor=blue
            ]{hyperref}
\makeatletter
\newcommand{\printfnsymbol}[1]{%
  \textsuperscript{\@fnsymbol{#1}}%
}
\makeatother
\begin{document}
%
% \title{Contribution Title\thanks{Supported by organization x.}}
\title{Click-aware Structure Transfer with Sample Weight Assignment for Post-Click Conversion Rate Estimation}
\titlerunning{Click-aware Structure Transfer with Sample Weight Assignment}
% If the paper title is too long for the running head, you can set
% an abbreviated paper title here
%

\author{Kai Ouyang\thanks{Equal contribution} \inst{1} \and
Wenhao Zheng\printfnsymbol{1} \inst{2}\orcidID{0000-0001-8227-8820} \and
Chen Tang\inst{1} \and
\\  Xuanji Xiao\inst{2}\orcidID{0000-0002-0499-8838} \Letter \and
Hai-Tao Zheng \inst{1,3} \Letter}
\authorrunning{Kai Ouyang et al.}

% % First names are abbreviated in the running head.
% % If there are more than two authors, 'et al.' is used.
% %
\institute{Tsinghua Universiy, China \\ 
\email{\{oyk20@mails, tc20@mails, zheng.haitao@sz\}.tsinghua.edu.cn} \\
\and Shopee Inc, Shenzhen, China. \\
\email{\{wenhao.zheng,charles.xiao\}@shopee.com} \\ \and
Pengcheng Laboratory, Shenzhen, China, 518055 \\
}
% \institute{Princeton University, Princeton NJ 08544, USA \and
% Springer Heidelberg, Tiergartenstr. 17, 69121 Heidelberg, Germany
% \email{lncs@springer.com}\\
% \url{http://www.springer.com/gp/computer-science/lncs} \and
% ABC Institute, Rupert-Karls-University Heidelberg, Heidelberg, Germany\\
% \email{\{abc,lncs\}@uni-heidelberg.de}}

\toctitle{Click-aware Structure Transfer with Sample Weight Assignment}
\tocauthor{Kai Ouyang et al.}
\maketitle              % typeset the header of the contribution
%

% \footnote{The first two authors contribute equally to this work.}
\footnote{This work was completed during Kai Ouyang's internship at Shopee, and Xuanji Xiao is the first corresponding author of this work.}

\begin{abstract}
Post-click Conversion Rate (CVR) prediction task plays an essential role in industrial applications, such as recommendation and advertising.
Conventional CVR methods typically suffer from the data sparsity problem as they rely only on samples where the user has clicked.  
To address this problem, researchers have introduced the method of multi-task learning, which utilizes non-clicked samples and shares feature representations of the Click-Through Rate (CTR) task with the CVR task.
However, it should be noted that the CVR and CTR tasks are fundamentally different and may even be contradictory.  
Therefore, introducing a large amount of CTR information without distinction may drown out valuable information related to CVR.  
This phenomenon is called the \textit{curse of knowledge} problem in this paper.
To tackle this issue, we argue that a trade-off should be achieved between the introduction of large amounts of auxiliary information and the protection of valuable information related to CVR.
Hence, we propose a \textbf{C}lick-aware \textbf{S}tructure \textbf{T}ransfer model with sample \textbf{W}eight \textbf{A}ssignment, abbreviated as \textbf{CSTWA}.
It pays more attention to the latent structure information, which could refine the input information related to CVR, instead of directly sharing feature representations.
Meanwhile, to capture the representation conflict between CTR and CVR, we calibrate the representation layer and reweight the discriminant  layer to excavate the click bias information from the CTR tower.
Moreover, it incorporates a sample weight assignment algorithm biased towards CVR modeling, to make the knowledge from CTR would not mislead the CVR.
Extensive experiments on industrial and public datasets have demonstrated that CSTWA significantly outperforms widely used and competitive models.


\end{abstract}

\section{Introduction}

The Conversion Rate (CVR) prediction task is crucial for ranking systems in modern industrial applications, such as e-commerce platforms and video platforms, as it is essential for better user experience and improving revenue.
Conventional CVR methods employ the same network architectures as Click-Through Rate (CTR) prediction task. 
However, in practice, the samples of the CVR task are usually much fewer than the CTR, with the former being only about 1\% of the latter.
The sparsity of training data makes CVR models suffer great difficulties in fitting, which leads to the well-known data sparsity problem of CVR~\cite{esmm_ma2018entire}.

Inspired by multi-task learning, recent CVR methods have introduced auxiliary tasks (\emph{e.g.}, the CTR task) to address the above challenge~\cite{esm2_wen2020entire,ple_tang2020progressive}.
They leverage samples of non-clicked impressions and even directly share the feature representations with auxiliary tasks.
Although the introduction of substantial external knowledge alleviates the data sparsity problem to a certain extent, this approach is not perfect.
On the one hand, the amount of knowledge introduced from auxiliary tasks (\emph{e.g.}, the CTR task) far exceeds that possessed by the CVR task, which reduces the model's ability to capture user conversion behavior. 
On the other hand, the introduced auxiliary knowledge and the knowledge related to CVR are not clearly distinguished, which makes the model easily assimilated by the introduced external knowledge.
These two reasons make it highly likely that the CVR model will be overwhelmed by a large amount of knowledge introduced (\emph{i.e.}, CTR knowledge).
This paper calls this phenomenon the \textit{curse of knowledge} problem.

However, the CVR task is inherently different from the CTR task. In some cases, they are even contradictory.
For example, item covers with sexy or erotic content tend to attract more clicks while few of them would lead to purchase.
Besides, although the number of clicks on some daily needs (\emph{e.g.}, kitchenware, toilet paper, \emph{etc}) of a user is usually low, the purchase rate is high.
Namely, there exists a huge gap between the information on the user's click behaviors (\emph{i.e.}, the CTR knowledge) and the information needed to model the user's actual conversion behaviors. 
Therefore, the curse of knowledge problem in the multi-task learning-based CVR models hinders their ability to capture real conversion behaviors, resulting in suboptimal performance.

To tackle this issue, we believe that it is necessary to alleviate the curse of knowledge problem that arises while introducing a large amount of auxiliary knowledge to solve the data sparsity problem.
Hence, we put forward a \textbf{C}lick-aware \textbf{S}tructure \textbf{T}ransfer model with sample \textbf{W}eight \textbf{A}ssignment algorithm, abbreviated as \textbf{CSTWA}, to protect the valuable knowledge related to CVR. 
We mine the task-independent information (\emph{i.e.}, latent structure information~\cite{mf_rendle2012bpr}) from the CTR and construct the Structure Migrator to inject it into the CVR feature representations.
Meanwhile, we design the Click Perceptron to model the click bias information, so that the CVR tower can capture the difference between clicked and non-clicked samples.
Moreover, we devise a novel sample weight assignment algorithm, named Curse Escaper.
It makes the model pay more attention to those samples whose CVR information and CTR information are contradictory.
In this way, we can further enhance the ability to model the users' real conversion behaviors and avoid the model being overwhelmed by CTR information.
We conduct extensive experiments and achieve state-of-the-art (SOTA) performance on industrial and public datasets, which demonstrates the superiority of our CSTWA.

To summarize, we mainly make the following fourfold contributions:
\begin{itemize}
    \item We elucidate the \textit{curse of knowledge} problem of the CVR methods based on multi-task learning.
    To alleviate this problem, we propose CSTWA, which can protect valuable information related to CVR while introducing auxiliary information.
    \item 
    We mine latent, task-independent \textit{item-item} and \textit{user-user} structure information from CTR, and we utilize it to filter the knowledge related to CVR. 
    \item We enhance the representation layer with a calibration technique and the discriminant layer with a brand-new sample weight assignment algorithm, which can explicitly mitigate the side effects of introducing knowledge of auxiliary tasks.
    \item Extensive experiments on industrial and public datasets demonstrate the superiority of CSTWA over competitive methods.
\end{itemize}

\section{Related Work}
\label{sec:rw}
This paper aims to alleviate the curse of knowledge problem in the CVR models based on multi-task learning.
Therefore, we briefly review the most related work from the following two aspects: (a) Conversion Rate Prediction, and (b) Multi-task Learning.

\subsection{Conversion Rate (CVR) Prediction}
Recommender systems are highly significant in contemporary society~\cite{ouyang2022social,ouyang2023mining}.
Conversion Rate (CVR) prediction is critical for the ranking systems of many industrial applications, such as recommendation systems~\cite{pnn_qu2016product,esmm_ma2018entire}, as it is directly related to user experience and final revenue.

Although research on CTR is prosperously developing~\cite{wdl_cheng2016wide,deepfm_guo2017deepfm,xdeepfm_lian2018xdeepfm,ffm_juan2016field,dcn_wang2017deep,autoint_song2019autoint,fibinet_huang2019fibinet}, there are few kinds of literature directly proposed for the CVR task~\cite{CVR_ls_yang2016large,CVR_mld_wen2019multi}.
Meanwhile, researchers often directly use the CTR prediction method to estimate the CVR, and only replace the training samples with the samples of clicked impressions.
However, this causes the well-known extremely data sparsity problem~\cite{esmm_ma2018entire}.

To address this challenge, recently proposed methods model CVR directly over the entire space, which includes the samples of clicked and non-clicked impressions.
For example, ESMM~\cite{esmm_ma2018entire} introduces two auxiliary tasks of predicting the post-view click-through rate (CTR) and post-view click-through \& conversion rate (CTCVR), instead of training the CVR model directly with samples of clicked impressions.
ESM$^2$~\cite{esm2_wen2020entire} models CVR prediction and auxiliary tasks simultaneously according to the conditional probability rule defined on the user behavior graph, in a multi-task learning framework.
Recently, AITM~\cite{aitm_xi2021modeling} proposes to adaptively learn what and how much information to transfer for different conversion stages of different users, achieving significantly better performance compared with previous methods.
Moreover, ESCCM$^2$~\cite{wang2022escm2} employs a counterfactual risk minimizer as a regularizer in ESMM~\cite{esmm_ma2018entire} to address both Inherent Estimation Bias (IEB) and Potential Independence Priority (PIP) issues simultaneously.

However, they either share feature representations with the CTR tower or directly use knowledge from CTR. 
They overlook or fail to address the curse of knowledge caused by introducing massive external auxiliary information, which limits their ultimate performance. 
In contrast, when we train the CVR model using samples from the entire space, we weaken the assimilation from auxiliary knowledge and enhance valuable CVR-related knowledge through structure transfer strategy and sample weight assignment algorithm.

\subsection{Multi-Task Learning}
To model the multi-stage nature of the user’s purchasing behavior (\emph{e.g.}, $impression \rightarrow click \rightarrow purchase$), and alleviate the model fitting difficulty caused by the data sparsity problem of the CVR task, the previous work tries to model the CVR task through a multi-task learning framework.
They model one or more auxiliary tasks for the CVR task (\emph{e.g.}, CTR task), expanding the sample space while capturing information about the multi-stage nature of the purchasing behavior~\cite{mlt_hadash2018rank,lu_ni2018perceive}.
ESMM series (ESMM~\cite{esmm_ma2018entire}, ESMM2~\cite{esm2_wen2020entire}) uses click signals (CTR) and post-click signals and uses shared feature-embedding layer to tackle the conversion sample sparsity.
MMOE series (MMOE\cite{mmoe_ma2018modeling}, SNR~\cite{snr_ma2019snr}, PLE~\cite{ple_tang2020progressive}) propose a multi-gate mixture of expert sub-networks, which provide a limited degree of sharing at the granularity of the sub-network.

These multi-task learning methods alleviate the data sparsity problem of the CVR task to a certain extent.
They capture the multi-stage nature of users' purchasing behavior, achieving considerable performance improvement.
However, they all ignore the fact that auxiliary tasks (\emph{e.g.}, CTR task) and the CVR task are essentially different and even contradictory.
Therefore, directly using the samples of auxiliary tasks or even directly sharing the feature representation can lead to the curse of knowledge problem, \emph{i.e.}, the CVR model will be overwhelmed by the knowledge of auxiliary tasks.

% 修改图片中的字体大小

\section{Method}
\label{sec:me}
In this section, we define the problem and the key notations, and introduce our proposed framework CSTWA in detail. 

\subsection{Problem Definition}
The task of Post-click Conversion Rate (CVR) prediction aims to predict the likelihood that a user will purchase an item, while the task of Click-Through Rate (CTR) prediction aims to predict the probability of a user clicking on an item. 
Both tasks utilize a feature set that includes user features, item features, and context features, most of which are categorical and can be represented using one-hot encoding.

We assume the observed dataset to be $\mathcal{S} = \{s_i = (x_i,y_i,z_i)\}|_{i=1}^{N}$, where $s_i=(x_i,y_i,z_i)$ denotes an impression sample, and $N$ is the total number of impression samples.
$x_i=\{f_1,\dots,f_n\}$ represents feature values of an impression sample, where $f$ denote a feature value.
The binary labels $y$ and $z$ indicate whether a click or conversion event occurs in the sample $s_i$, respectively, and can take on the values of 1 or 0.

The estimation of click-through rate (CTR) and conversion rate (CVR) involves predicting the probabilities of the following events, respectively:
\begin{equation}
\begin{split}
& \mathcal{P}_{\text{CTR}} = \mathcal{P}(y=1|x),\ \mathcal{P}_{\text{CVR}} = \mathcal{P}(z=1|x).
\end{split}
\end{equation}

In this paper, we define $X^{U}=\{x^{U}_{1},\dots,x^{U}_{N^{U}}\}$ and $X^{I}=\{x^{I}_{1},\dots,x^{I}_{N^{I}}\}$ as sets representing all users and items, respectively. Here, $x^{U}$ and $x^{I}$ represent all feature values of a user or item (identified by their unique ID), where $|x^{U}|=l^{U}$ and $|x^{I}|=l^{I}$ denote the number of features for each user or item, and $N^{U}$ and $N^{I}$ denote the total number of users and items, respectively.

\begin{figure*}[t]
\centering
\includegraphics[width=1\textwidth]{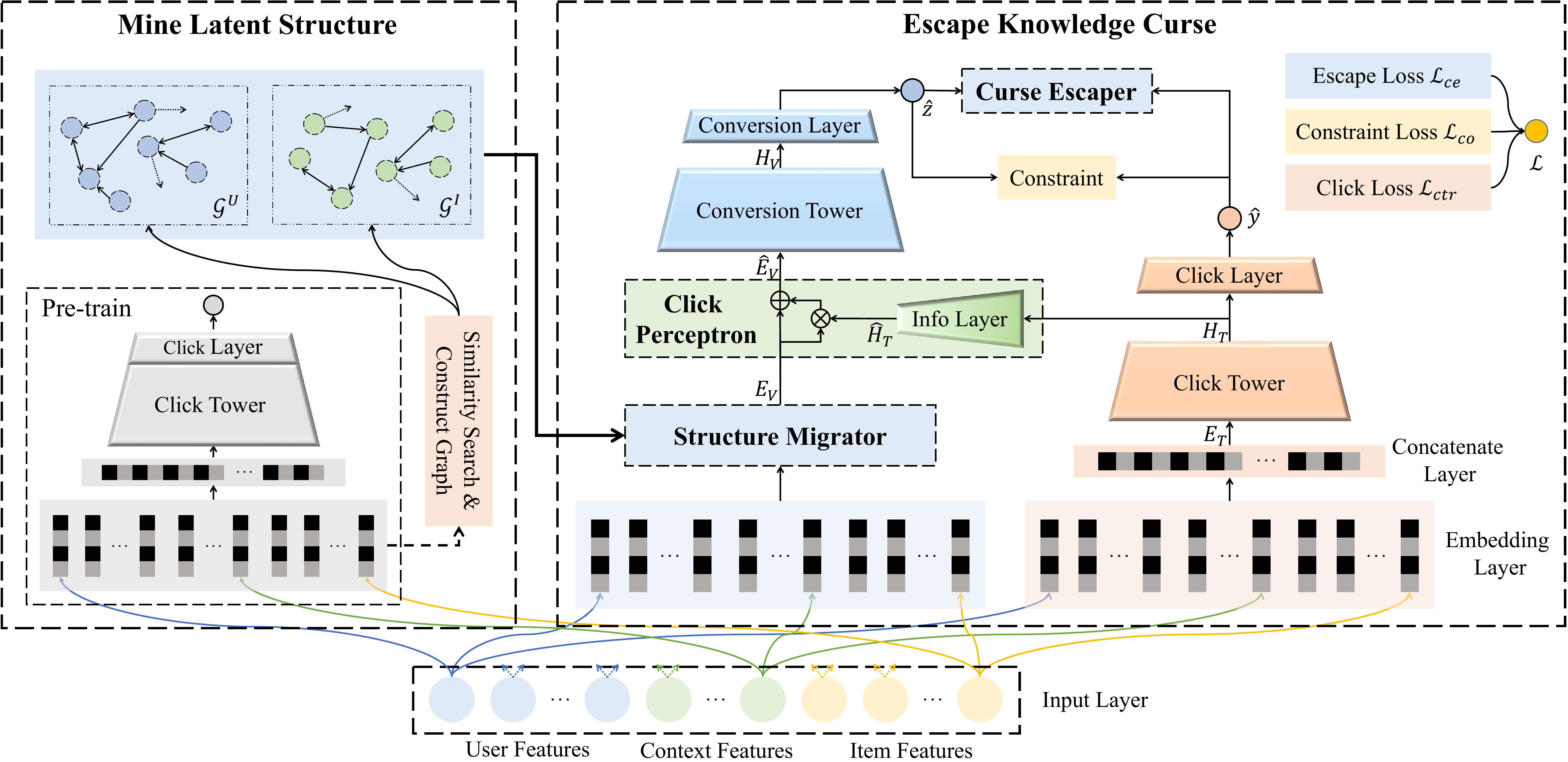} 
\caption{The architecture of CSTWA. It could be divided into two parts. The first part is the Mine Latent Structure. The second one is the Escape Knowledge Curse, which contains three important components: Structure Migrator, Click Perceptron, and Curse Escapter.}
\label{fig:architecture}
\vspace{-8pt}
\end{figure*}

\subsection{Motivation and Architecture}
\label{chapter:3.2}
As the user's purchase behavior only occurs after impressions and clicks, sufficient interactions are often needed before a purchase is made, resulting in limited training samples for CVR.
This poses a serious data sparsity problem for CVR. 
Although utilizing knowledge of CTR can mitigate this issue, it introduces the curse of knowledge where the knowledge of CVR is overshadowed by a vast amount of CTR knowledge.

To address this problem, we propose that we should learn latent and task-independent structure information instead of directly using the feature representations of CTR. 
We also model the click bias information of samples to enable the model to differentiate between clicked and non-clicked samples. 
Furthermore, we introduce a brand-new weight assignment algorithm called Curse Escaper, which emphasizes the role of false negative samples (\emph{i.e.}, the output of CTR tower is high, but $z=0$) and false positive samples (\emph{i.e.}, the output of CTR tower is low, but $z=1$) in the training process.
The above three components enable our model to escape the curse of CTR knowledge.

The architecture of CSTWA is depicted in Figure~\ref{fig:architecture}. The entire framework comprises the following components, which we will elaborate on later in this section: 
\begin{itemize}
    \item Mine Latent Structure, which mines latent pairwise \textit{user-user} and \textit{item-item} structure information from the feature representations of the pre-trained CTR model. It generates user and item graphs;
    \item Escape Knowledge Curse, which relies on three components, namely, (i) Structure Migrator, (ii) Click Perceptron, and (iii) Curse Escaper, to model the CVR task and alleviate the curse of knowledge problem.
\end{itemize}

\subsection{Mine Latent Structure}
Previous research has shown that users are more likely to interact with items that are similar to those they have interacted with before, rather than different ones~\cite{mf_rendle2012bpr}. 
This collaborative filtering relationship, also known as structure information, is independent of the task. 
Therefore, the collaborative filtering knowledge discovered on the CTR task can be applied to CVR modeling as additional knowledge.

\subsubsection{Pre-train.}
To capture the collaborative filtering knowledge, we pre-trained a simple CTR model that only contained a Multi-Layer Perceptron (MLP) layer. 
In this study, we utilized the basic MLP as the backbone structure for the CVR, CTR, and pre-trained CTR models, which is similar to previous research~\cite{aitm_xi2021modeling}.

The pre-trained CTR model's structure is consistent with the rightmost part of Figure~\ref{fig:architecture}, and it consists of five components: 
(a) Input Layer, where samples are input.
(b) Embedding Layer, which transforms the sparse and dense features into feature embedding.
(c) Concatenate Layer, which concatenates the feature embeddings.
(d) Click Tower, which contains a multi-layer perceptron, the Dropout, and the ReLU activation function.
(e) Click Layer, which consists of a multi-layer perceptron and the Sigmoid activation function.

We train this CTR model using impression samples and extract its Embedding Layer to capture collaborative filtering knowledge in the CTR task.

\subsubsection{Similarity Search.}
To derive collaborative filtering insights from the CTR task, we compute similarities between items and between users.
To begin, we concatenate the feature embeddings of a user according to their ID to generate the user representation. 
Similarly, we employ the same procedure for items to derive their representation:
\begin{equation}
\begin{split}
r^{U}_{\text{pre}}=[e^{U}_{1,\text{pre}}||\dots||e^{U}_{l^{U},\text{pre}}], 
r^{I}_{\text{pre}}=[e^{I}_{1,\text{pre}}||\dots||e^{I}_{l^{I},\text{pre}}],
\end{split}
\label{eq:SS1}
\end{equation}
where $e_{\text{pre}} \in \mathbb{R}^{d}$ denotes the feature embedding obtained from the pre-trained CTR model, $d$ is the embedding size, and $||$ denotes the concatenate operation.
Besides, $r^{U}_{\text{pre}} \in \mathbb{R}^{d^{U}}, r^{I}_{\text{pre}} \in \mathbb{R}^{d^{I}}$ represent the user/item representation, where $d^{U}=l^{U}\times d$, $d^{I}=l^{I}\times d$, and 
$l^{U}, l^{I}$ denote the number of features for the user and item, respectively.

To evaluate the similarity between users, we opt for the straightforward and non-parametric cosine similarity measure~\cite{amgcn_wang2020gcn}:
\begin{equation}
    S^{U}_{ij} = \frac{(r^{U}_{i,\text{pre}})^{T}r^{U}_{j,\text{pre}}}{||r^{U}_{i,\text{pre}}||\ ||r^{U}_{j,\text{pre}}||},
\label{eq:SS2}
\end{equation}
where $S^{U}_{ij}$ is scalar, which represents the semantic correlations between two users $i,j$. $S^{U} \in \mathbb{R}^{N^{U}\times N^{U}}$ denotes the similarity matrix of users.
Similar to Eqs.~\ref{eq:SS1} and \ref{eq:SS2}, we can also obtain the similarity matrix $S^{I}$ for items.

To reduce the time cost of constructing the similarity matrix, we use the Faiss \footnote{\url{https://github.com/facebookresearch/faiss}}~\cite{faiss_johnson2019billion} library, which can efficiently performs similarity searches.

\subsubsection{Construct Graph.}
To visualize these collaborative filtering relationships, we construct graphs according to the similarity matrix.
Typically, the adjacency matrix of a graph should contain non-negative values, but in our case, the values of $S_{ij}$ fall within the range of $[-1,1]$.
Hence, we set the negative terms of $S$ to zeros.
Additionally, we recognize that fully-connected graphs are computationally intensive and are likely to introduce noisy and unimportant edges~\cite{noisy_chen2020iterative}, so we aim to create sparse graphs.

Specifically, we only retain edges with the top-$K$ confidence scores for each node $i$, resulting in the sparsified adjacency matrix $G^{U}$:
\begin{equation}
G^{U}_{ij}=
\left\{
\begin{array}{rcl}
S^{U}_{ij},      &      & \text{if} \ \ S^{U}_{ij} \in \text{top-}K(S^{U}_{i,:}), \\
0,           &      & \text{otherwise},
\end{array} 
\right. \\
\label{eq:GC1}
\end{equation}
where $S^{U}_{i,:}$ denotes the $i$-row of $S^{U}$, and $G^{U}$ is the sparsified adjacency matrix.

To alleviate the exploding or vanishing gradient problem~\cite{normalize_kipf2016semi}, we normalize the adjacency matrix as:
\begin{equation}
    \mathcal{G}^{U} = (D)^{-\frac{1}{2}}G^{U}(D)^{-\frac{1}{2}},
\label{eq:GC2}
\end{equation}
where $D \in \mathbb{R}^{d^{U}}$ is the diagonal degree matrix of ${G}^{U}$, $\mathcal{G}^{U}$ denotes the final graph for users.
Similar to Eqs.~\ref{eq:GC1} and \ref{eq:GC2}, we can also obtain the sparsified adjacency matrix ${G}^{I}$ and the final graph $\mathcal{G}^{I}$ for items.
To reduce the storage cost of the adjacency matrix, we use the Sparse\footnote{\url{https://pytorch.org/docs/stable/sparse.html}} library of Pytorch~\cite{pytorch_paszke2017automatic}.

\subsection{Escape Knowledge Curse}
After obtaining the collaborative filtering information based on the CTR task (\emph{i.e.}, $\mathcal{G}^{U}$ and $\mathcal{G}^{I}$), we implement the three components discussed in Chapter \ref{chapter:3.2} to alleviate the problem of the curse of knowledge.
To illustrate the method's overall flow, we outline the transformation process of a sample $s=(x,y,z)$ from input to output. 

To begin with, we must convert the characteristics of the input sample into a feature representation through an embedding lookup. 
Similar to previous CVR methods that utilize multi-task learning, we train the CVR model on the entire impression space and utilize the CTR task as an auxiliary. 
However, we differ in that we have two separate Embedding Layers for the CVR and CTR tasks, which we refer to as $V$ and $T$, respectively. 
Hence, the process of embedding lookup can be described as follows:
\begin{equation}
\begin{split}
    & e_{i,V} = V(f_i),\ e_{i,T} = T(f_i), \\
\end{split}
\end{equation}
where $e_i \in \mathbb{R}^{d}$ denotes the feature embedding for $i$-th feature $f_i$ of the sample $s$, and $d$ is the embedding size.

Then, we can obtain the representation of the sample $s$ for the CTR tower as follows:
\begin{equation}
\begin{split}
    E_{T}  = [r_{s,T}^{U}||r_{s,T}^{I}||r_{s,T}^{C}] = [e_{1,T}^{U} || \dots || e_{l^{U},T}^{U} || e_{1,T}^{I} || \dots || e_{l^{I},T}^{I} || e_{1,T}^{C} || \dots || e_{l^{C},T}^{C} ], \\
\end{split}
\end{equation}
where $E_{T} \in \mathbb{R}^{d^{U}+d^{I}+d^{C}}$ is the input of CTR tower,  $r_{s,T}^{U} \in \mathbb{R}^{d^{U}}$ is the user-side feature representation of the sample $s$, $\ r_{s,T}^{I} \in \mathbb{R}^{d^{I}}$ is the item-side feature representation, $\ r_{s,T}^{C} \in \mathbb{R}^{d^{C}}$ denotes the context feature representation. The value of $d^{C}$ is calculated as $l^{C}\times d$, where $l^{C}$ is the number of context features present in sample $s$.

\subsubsection{Structure Migrator (SM).}
In this part, we migrate the task-independent structure information (\emph{i.e.}, $\mathcal{G}^{U}$ and $\mathcal{G}^{I}$) mined from the pre-trained CTR model to the CVR model.

We repeat the Eq.~\ref{eq:SS1} on the Embedding Layer $V$ of the CVR model to obtain the user/item representation matrices:
\begin{equation}
\begin{split}
& R^{U}=\{r^{U}_{1,V},\dots,r^{U}_{N^{U},V}\},\ R^{I}=\{r^{I}_{1,V},\dots,r^{I}_{N^{I},V}\}, \\
\end{split}
\label{eq:SM1}
\end{equation}
where $R^{U} \in \mathbb{R}^{d^{U}\times N^{U}}$, $R^{I} \in \mathbb{R}^{d^{I}\times N^{I}}$, $r^{U} \in \mathbb{R}^{d^{U}}$, $r^{I} \in \mathbb{R}^{d^{I}}$, and $N_U, N_I$ denote the total number of users and items, respectively.

Next, we utilize a straightforward yet efficient message propagation and aggregation process, which does not require any feature transformation or activation.
This process is computationally efficient as well.
The $l$-th layer of this process can be expressed as:
\begin{equation}
    {R}^{U}_{(l)} = \mathcal{G}^{U} {R}^{U}_{(l-1)},
\label{eq:SM2}
\end{equation}
where ${R}^{U}_{(l)} \in \mathbb{R}^{d^{U}\times N_{U}}$ is the $l$-th layer embedding matrix, and the initial ${R}^{U}_{(0)}$ is $R^{U}$.
After stacking $L$ layers, $R^{U}_{(L)}$ encodes the high-order user-user relationships. 

Similar to Eqs.~\ref{eq:SM1} and \ref{eq:SM2}, we can also obtain the item's $R^{I}{(L)}$. 
As a result, we obtain new user/item embeddings, \emph{i.e.}, $R^{U}{(L)}$ and $R^{I}{(L)}$, which encode collaborative filtering knowledge from CTR. 
To inject the collaborative filtering knowledge into CVR modeling more smoothly, we set a hyper-parameter to control the update of CVR features. 
In the $t$-th epoch ($t>1$), the update of $R^{U}_{(L)}$ can be formulated as:
\begin{equation}
    R^{U}_{(L,t)} = \alpha R^{U}_{(L,t)} + (1 - \alpha) R^{U}_{(L,t-1)},
\end{equation}
where $\alpha$ controls the proportion of updates. Similarly, we update $R^{I}_{(L,t)}$ through the same process.

We then generate the representation of the sample $s$ for the CVR tower:
\begin{equation}
\begin{split}
    E_{V} = [r_{s,V}^{U}||r_{s,V}^{I}||r_{s,V}^{C}] = [R^{U}_{(L,t)}[s]||R^{I}_{(L,t)}[s]||e_{1,V}^{C} || \dots || e_{l^{C},V}^{C}], \\
\end{split}
\end{equation}
where $E_{V} \in \mathbb{R}^{d^{U}+d^{I}+d^{C}}$ is the input of CVR tower, and $R^{\cdot}_{(L,t)}[\cdot]$ means searching for the embeddings corresponding to the user or item in the sample $s$ from the embedding matrix $R^{U}_{(L,t)}$ or $R^{I}_{(L,t)}$.

\subsubsection{Click Perceptron (CP).}
To distinguish between clicked impression samples and non-clicked impression samples during training, we incorporate the hidden representation of the CTR auxiliary task into CVR modeling as bias information for the samples.

To capture CTR knowledge in the sample, we employ Click Tower on $E_{T}$:
\begin{equation}
   {H}_{T} = \text{TW}_{\text{click}}(E_{T}), \\
\end{equation}
where ${H}_{T} \in \mathcal{R}^{d^o}$, $d^o$ is the embedding size of the output layer of the Click tower, and $\text{TW}_{\text{click}}$ means the Click Tower.

Next, we construct the Info Layer to incorporate bias information while preserving the original distribution and mitigating the instability of model training. 
This layer comprises a multi-layer perceptron and the Sigmoid activation function:
\begin{equation}
    \hat{H}_{T} = \text{Sigmoid}(\text{MLP}(H_{T})), \\
\end{equation}
where $\hat{H}_{T} \in \mathbb{R}^{d^{U}+d^{I}+d^{C}}$, which is composed of a series of numbers between $[0,1]$.
MLP represents the multi-layer perceptron.
We multiply $\hat{H}_{T}$ by 2 to keep its mean around 1 and inject it as bias information into the modeling of CVR:
\begin{equation}
   \hat{E}_{V} = E_{V}(1 + 2 \hat{H}_{T}), \\
\end{equation}
where $\hat{E}_{V} \in \mathbb{R}^{d^{U}+d^{I}+d^{C}}$ denotes the hidden representation of sample $s$ in CVR modeling, which is injected with collaborative filtering knowledge and click bias information.

To extract CVR knowledge and the information introduced externally, we employ Conversion Tower on $\hat{E}_{V}$:
\begin{equation}
\begin{split}
   H_{V} = \text{TW}_{\text{conv}}(\hat{E}_{V})), \\
\end{split}
\end{equation}
where ${H}_{V} \in \mathcal{R}^{d^o}$, and $\text{TW}_{\text{conv}}$ means the Conversion Tower.

To estimate the probability of user click and purchase respectively, we construct the Conversion Layer and Click Layer, which are comprised of the multi-layer perceptron and Sigmoid activation function:
\begin{equation}
\begin{split}
    \hat{y} = \text{Sigmoid}(\text{MLP}(H_{T})), \hat{z} = \text{Sigmoid}(\text{MLP}(H_{V})), \\ 
\end{split}
\end{equation}
where $\hat{y},\hat{z}$ are the predicted probabilities of click and purchase behavior occurring in the sample $s$, respectively.

\subsubsection{Curse Escaper (CE).}
To further strengthen the model's ability to model users' real conversion behavior, we add a brand-new weight assignment algorithm, named Curse Escaper, into the loss function.
The loss function of CSTWA is mainly composed of three parts, as follows.

First, for the CTR task, we minimize the cross-entropy loss of it:
\begin{equation}
    \mathcal{L}_{\text{ctr}}=-\frac{1}{N}\sum_{i=1}^{N}(y_i \log \hat{y}_i+(1-y_i)\log(1-\hat{y}_i)),
\end{equation}
where $N$ is the total number of samples in the entire sample space $\mathcal{S}$, $y_i$ is the click label of the $i$-th sample, and $\hat{y}_i$ is the predicted click probability.

To alleviate the curse of knowledge problem, we increase the weights of the false positive samples (\emph{i.e.}, samples where the CTR predictive value is high but no conversion behavior actually occurs) and the false negative samples (\emph{i.e.}, samples where the CTR predictive value is low but conversion behavior actually occurs) in the loss function.

Technically, we predefine two hyperparameters, $pos$ and $neg$, which indicate the thresholds for high CTR predictive value and low CTR predictive value, respectively.
Mathematically, the loss function of CVR can be formulated as:
\begin{equation}
\begin{split}
& \mathcal{L}_{\text{ce}}=-\frac{1}{N}\sum_{i=1}^{N}(\mathcal{A} z_i \log \hat{z}_i+ \mathcal{B} (1-z_i)\log(1-\hat{z}_i)), \\
& \text{where},\ \mathcal{A} = \text{max}(1,(\frac{neg}{\hat{y}_i})^{\gamma}),\ \mathcal{B} = \text{max}(1,(\frac{\hat{y}_i}{pos})^{\gamma}), \\
\end{split}
\end{equation}
where $\hat{y}_i$ is the CTR predictive value, $z_i$ is the conversion label of the $i$-th sample, and $\gamma$ controls the magnitude of weight enhancement.
To prevent excessive weighting, we set upper limits for $\mathcal{A}$ and $\mathcal{B}$. During the experiment, we searched for the optimal values between 2 and 10, and ultimately restricted $\mathcal{A}$ and $\mathcal{B}$ to be below 4.

Besides, considering that the purchase behavior occurs after the click, we follow AITM~\cite{aitm_xi2021modeling} to set an extra loss function to constrain the CVR predictive value to be less than the CTR predictive value:
\begin{equation}
    \mathcal{L}_{\text{co}} = \frac{1}{N}\sum_{i=1}^{N}\text{max}(\hat{z}_i-\hat{y}_i, 0).
\end{equation}
It outputs the positive penalty term only when $\hat{z}_i > \hat{y}_i$, otherwise, it outputs 0.
Finally, the loss function of our method can be formulated as:
\begin{equation}
    \mathcal{L} = w_{1}\mathcal{L}_{\text{ctr}} + w_{2}\mathcal{L}_{\text{ce}} + w_{3} \mathcal{L}_{\text{co}},
\end{equation}
where $w_{1}, w_{2}, w_{3}$ are weights of $\mathcal{L}_{\text{ctr}},\mathcal{L}_{\text{ce}},\mathcal{L}_{\text{co}}$, which are set to (1, 1, 0.6) in this paper, respectively

\section{Experiments}
In this section, we perform experiments to evaluate the proposed framework against various baselines on industrial and public real-world datasets, and answer the following Research Questions (RQs):
\begin{itemize}
    \item \textbf{RQ1:} How does our method perform compared with the baseline models on the public datasets?
    \item \textbf{RQ2:} How do the three key components affect the final performance?
    \item \textbf{RQ3:} How sensitive our model is to perturbations of several key hyper-parameters?
\end{itemize}

\subsection{Experimental Settings}
In this part, we introduce the benchmark dataset, the evaluation metrics, the state-of-the-art methods involved in the comparison, and the implementation details.

\begin{table}[t]
\caption{Statistics of the public datasets after processing, where M means million, and “\%Positive” means the percentage of clicked/converted samples in the train set.}
	\begin{center}
 
	\resizebox{0.6\textwidth}{!}{
 
		\begin{tabular}{c|ccc}
            \toprule
			Dataset & \#Users/\#Items & \#Train/\#Validation/\#Test & \%Positive(\%) \\ 
			\midrule
			Industrial & 30M/ 0.28M & 465M/45M/44M &  7.04/0.28 \\
            Public & 0.2M/0.5M & 38M/4.2M/43M &  3.89/0.02\\
			\bottomrule
		\end{tabular}
    }
	\end{center}
	\label{tb:dataset}
 \vspace{-15pt}
\end{table}

\subsubsection{Datasets.}
We conduct extensive experiments on the following two datasets: \textit{Industrial dataset}: The industrial dataset contains all samples of the livestream platform of our App, which is one of the largest e-commerce platform in the world, in a continuous period of time.
We divide the training set, validation set and test set in chronological order. 
\textit{Public dataset}: The public dataset is the Ali-CCP (Alibaba Click and Conversion Prediction) \cite{esmm_ma2018entire} dataset \footnote{\url{https://tianchi.aliyun.com/dataset/dataDetail?dataId=408}}. We follow the previous work to use all the single-valued categorical features and randomly take 10\% of the train set as the validation set to verify the convergence of all models.
We follow the previous work \cite{aitm_xi2021modeling} to filter the features whose frequency less than 10. 
The statistics of datasets are shown in Table \ref{tb:dataset}.

\subsubsection{Evaluation Metrics.}
In the offline experiments, to comprehensively evaluate the effectiveness of our method and compare it with the baseline methods, we follow the existing
works~\cite{esmm_ma2018entire,ple_tang2020progressive,aitm_xi2021modeling} to adopt the standard metric Area Under Curve (AUC), which is widely used in the recommendation and advertising systems and can reflect the ranking ability.
The mean and standard deviation (std) is reported over five runs with different random seeds.
We report the AUC on the auxiliary task and the focused main task (\emph{i.e.}, click and purchase prediction task).

\subsubsection{Baseline Methods.}
We compare the proposed method with the following representative and mainstream models: 
\textbf{MLP}: It is the base structure of our framework, which consists of the Click/Conversion Tower and Click/Conversion Layer. 
\textbf{ESMM}~\cite{esmm_ma2018entire,esm2_wen2020entire}: ESMM and ESM$^2$ use a unified multi-task learning framework to predict purchase and post-click behaviors over the entire space to relieve the sample selection bias problem.
\textbf{OMoE}~\cite{mmoe_ma2018modeling}: The Expert-Bottom pattern in OMoE (One-gate Mixture-of-Experts) incorporates experts by utilizing a single gate shared across all tasks.
\textbf{MMoE}~\cite{mmoe_ma2018modeling}: It is designed to integrate experts via multiple gates in the Gate Control. Moreover, it explicitly learns to model task relationships from data.
\textbf{PLE}~\cite{ple_tang2020progressive}: Progressive Layered Extraction (PLE) with Expert-Bottom pattern separates task-shared experts and task-specific experts explicitly.
\textbf{AITM}~\cite{aitm_xi2021modeling}: It is a contemporaneous work that proposes the attention-based AIT module which can adaptively learn what and how much information to transfer for different stages of different audiences.
\textbf{ESCM$^2$}~\cite{wang2022escm2}: It is devised to augment ESMM with counterfactual regularization. This is the state-of-the-art method in the CVR prediction task.

To be fair, all models, including CSTWA, utilize the same fundamental network structure and hyper-parameters in their Multi-Layer Perceptron.

\subsubsection{Implementation Details.}
We implemented our method using PyTorch~\cite{pytorch_paszke2017automatic}. To ensure a fair comparison, we set the embedding dimension $d$ to 5 for all models. We used the Adam optimizer with a learning rate of 0.001, a batch size of 2000, and 10 epochs. L2 regularization is set to 1e-6.
In the Expert-Bottom pattern, the dimensions of layers in the MLP-Expert are set to $[64,32,16]$.
In the Probability-Transfer pattern, the dimensions of layers in the MLP-Tower (\emph{i.e.}, Click/Conversion Tower) are set to $[128,64,32]$.
That is, $d^o = 32$.
The dropout rates in each layer are set to $[0.1,0.3,0.3]$, and the activation function used is ReLU.
In addition, we perform a grid search on the validation set to find the optimal hyper-parameters. We tune the $K$ parameter of the top-$K$ function using values of $4, 8, 16,$ and $32$, ultimately setting it to 8.
Meanwhile, we set $L$ to 1. 
The values for $\alpha$ and $\gamma$ were set to 0.3 and 3, respectively. 
For the hyper-parameters $pos$ and $neg$, we set them to the 99-th and 10-th percentile of the previous 10,000 CTR predicted values, respectively.
It is worth mentioning that, in the industrial dataset, we remove the Structure Migrator due to time cost considerations.

\begin{table*}[t]
\centering
\caption{The AUC performance (mean$\pm $std) on the industrial and public datasets. The Gain means the mean AUC improvement compared with the MLP. 
We bold the best results and indicate the second-best results with an underline.  “$^{\ast}$” indicates that the improvement is statistically significant at $p$-value $<$ 0.05 over paired samples t-test.}
\resizebox{1\textwidth}{!}{
\begin{tabular}{c|cccc|cccc}
\toprule
\textbf{Dataset} & \multicolumn{4}{c}{\textbf{Industrial Dataset}}&\multicolumn{4}{c}{\textbf{Public Dataset}} \\
\cmidrule(lr){2-5}\cmidrule(lr){6-9}
\textbf{Model} & \textbf{Click AUC} & \textbf{Gain} & \textbf{Purchase AUC} & \textbf{Gain}  & \textbf{Click AUC} & \textbf{Gain} & \textbf{Purchase AUC} & \textbf{Gain} \\
\midrule
MLP & 0.7970$\pm $0.0020 & \_ & 0.8466$\pm $0.005 & \_ & 0.6147$\pm $0.0012 & \_ & 0.5789$\pm $0.0042 & \_  \\
\midrule
ESMM & 0.7971$\pm $0.0012 & +0.0001 & 0.8579$\pm $0.0031 & +0.0113 & 0.6152$\pm $0.0023 & +0.0005 & 0.6376$\pm $0.0050 & +0.0587  \\
OMoE & 0.7969$\pm $0.0012 & -0.0001 & 0.8520$\pm $0.0040 & +0.0054 & \textbf{0.6192$\pm $0.0022} & +0.0045 & 0.6412$\pm $0.0061 & +0.0623  \\
MMoE & \underline{0.7982$\pm $0.0013} & +0.0012 & 0.8503$\pm $0.0032 & +0.0037 & 0.6170$\pm $0.0021 & +0.0023 & 0.6439$\pm $0.0035 & +0.0650  \\
PLE & \textbf{0.7989$\pm $0.0012} & +0.0019 & 0.8516$\pm $0.0022 & +0.0050 & 0.6166$\pm $0.0015 & +0.0019 & 0.6446$\pm $0.0026 & +0.0657  \\
AITM & 0.7884$\pm $0.0014 & -0.0086 & \underline{0.8587$\pm $0.0033} & +0.0121 & \underline{0.6183$\pm $0.0016} & +0.0036 & \underline{0.6456$\pm $0.0047} & +0.0667  \\
ESCM$^2$ & 0.7973$\pm $0.0019 & +0.0003 & 0.8505$\pm $0.0050 & +0.0039 & 0.6176$\pm $0.0012 & +0.0029 & 0.6427$\pm $0.0041 & +0.0638  \\
\midrule
CSTWA & 0.7938$\pm $0.0014 & -0.0032 & \textbf{0.8613$\pm $0.0024}$^{\ast}$ & +0.0147 & 0.6160$\pm $0.0020 & +0.0013 & \textbf{0.6532$\pm $0.0028}$^{\ast}$ & +0.0743  \\
\bottomrule
\end{tabular}
}
\label{tb:results}
\vspace{-8pt}
\end{table*}

\subsection{Experimental Results}
In this part, we report our experimental results and conduct a detailed analysis to investigate CSTWA framework.

\subsubsection{Main Results (RQ1).}
We report the AUC scores of all models on the offline test set. 
The results of purchase AUC are shown in Table~\ref{tb:results}, and we can draw the following insightful observations:

(a). Compared to the single-task model MLP, multi-task models generally exhibit better performance. This indicates that achieving satisfactory performance for the single-task model is difficult due to the extreme data sparsity. In fact, if the CVR model is trained solely with samples of clicked impressions, many of the feature embeddings may not be trained sufficiently. 

(b). ESMM and ESCM$^2$ models based on the Probability-Transfer pattern improve slightly as they only consider simple probability information transfer between adjacent tasks. On the other hand, the models based on the Expert-Bottom pattern, regulate the shared information among tasks and perform better than many baselines.

(c). AITM enhances CVR estimation by explicitly modeling inter-task relationships, while PLE outperforms baselines by separating shared and specific task experts. This demonstrates that explicitly modeling the transfer of information between tasks can improve performance.

(d). All existing works overlook the curse of knowledge caused by too much auxiliary task information. As a result, CSTWA is the best in CVR prediction among various baselines. This shows that extracting and transferring CTR knowledge, adding sample bias information, and enhancing CVR information can boost the main task.

\begin{table}[t]
\centering
\caption{The AUC performance (mean$\pm $std) of the variants on the public dataset. }
\resizebox{0.6\textwidth}{!}{
\begin{tabular}{c|cccc}
\toprule
\textbf{Dataset} & \multicolumn{4}{c}{\textbf{Public Dataset}} \\
\cmidrule(lr){2-5}
\textbf{Model} & \textbf{Click AUC} & \textbf{Gain} &  \textbf{Purchase AUC} & \textbf{Gain}  \\
\midrule
MLP & 0.6147$\pm $0.0012  & \_ & 0.5789$\pm $0.0042  & \_ \\
\midrule
$+$ SM & \textbf{0.6207$\pm $0.0031} & +0.0060 & 0.6372$\pm $0.0022 & +0.0583 \\
$+$ CP & 0.6159$\pm $0.0008  & +0.0012 & 0.6357$\pm $0.0034  & +0.0568   \\
$+$ CE & 0.6142$\pm $0.0010  & -0.0005 & 0.6312$\pm $0.0024  & +0.0523  \\
$+$ CP,\ CE & 0.6151$\pm $0.0022  & +0.004 & 0.6423$\pm $0.0025  & +0.0634 \\
$+$ SM,\ CE & \underline{0.6162$\pm $0.0012}  & +0.0015 & 0.6422$\pm $0.0027  & +0.0633 \\
$+$ SM,\ CP & 0.6154$\pm $0.0016  & +0.0007 & \underline{0.6503$\pm $0.0016}  & +0.0714 \\
\midrule
CSTWA & 0.6160$\pm $0.0020  & +0.0013 & \textbf{0.6532$\pm $0.0028}$^{\ast}$ & +0.0743 \\
\bottomrule
\end{tabular}
}
\label{tb:as}
\vspace{-8pt}
\end{table}

\subsubsection{Ablation Study (RQ2).}
To investigate the importance of critical components in CSTWA, namely the Structure Migrator (SM), Click Perceptron (CP), and Curse Escaper (CE), we conduct extensive ablation experiments to examine how they affect the final performance. 
The experimental results are presented in Table~\ref{tb:as}.

When utilizing any of the three key components, there is a significant increase in performance. 
This demonstrates that introducing bias information to enable the model to capture stage information in the sample can further improve performance. 
Additionally, explicitly enhancing the model's understanding of conversion behavior through the use of weight enhancement algorithms in the loss function can alleviate the curse of knowledge problem. 
Furthermore, when utilizing the SM component, there is a noticeable performance increase, indicating that task-independent structure information is effective.
The use of the CE component leads to a decrease in CTR-related metrics, as we explicitly enhance the weight of CVR-related knowledge in the samples. 
However, this decrease is acceptable because our main task is to model the CVR.
\begin{figure}[t]
\centering
	\subfloat[The effectiveness of $L$ and $\gamma$ \label{fig:a}]{
		\includegraphics[scale=0.18]{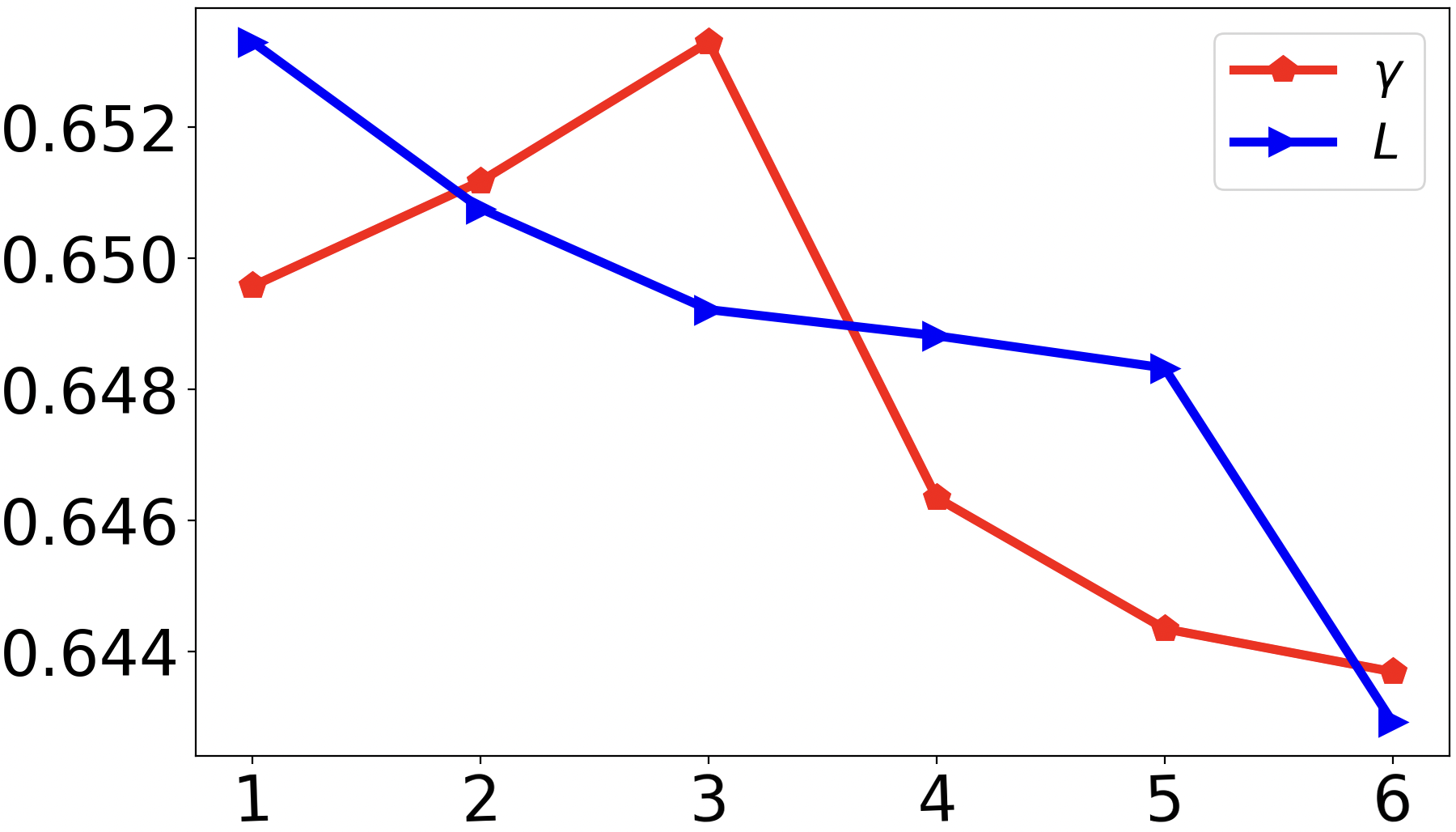}}
	\subfloat[The effectiveness of $\alpha$ \label{fig:b}]{
		\includegraphics[scale=0.09]{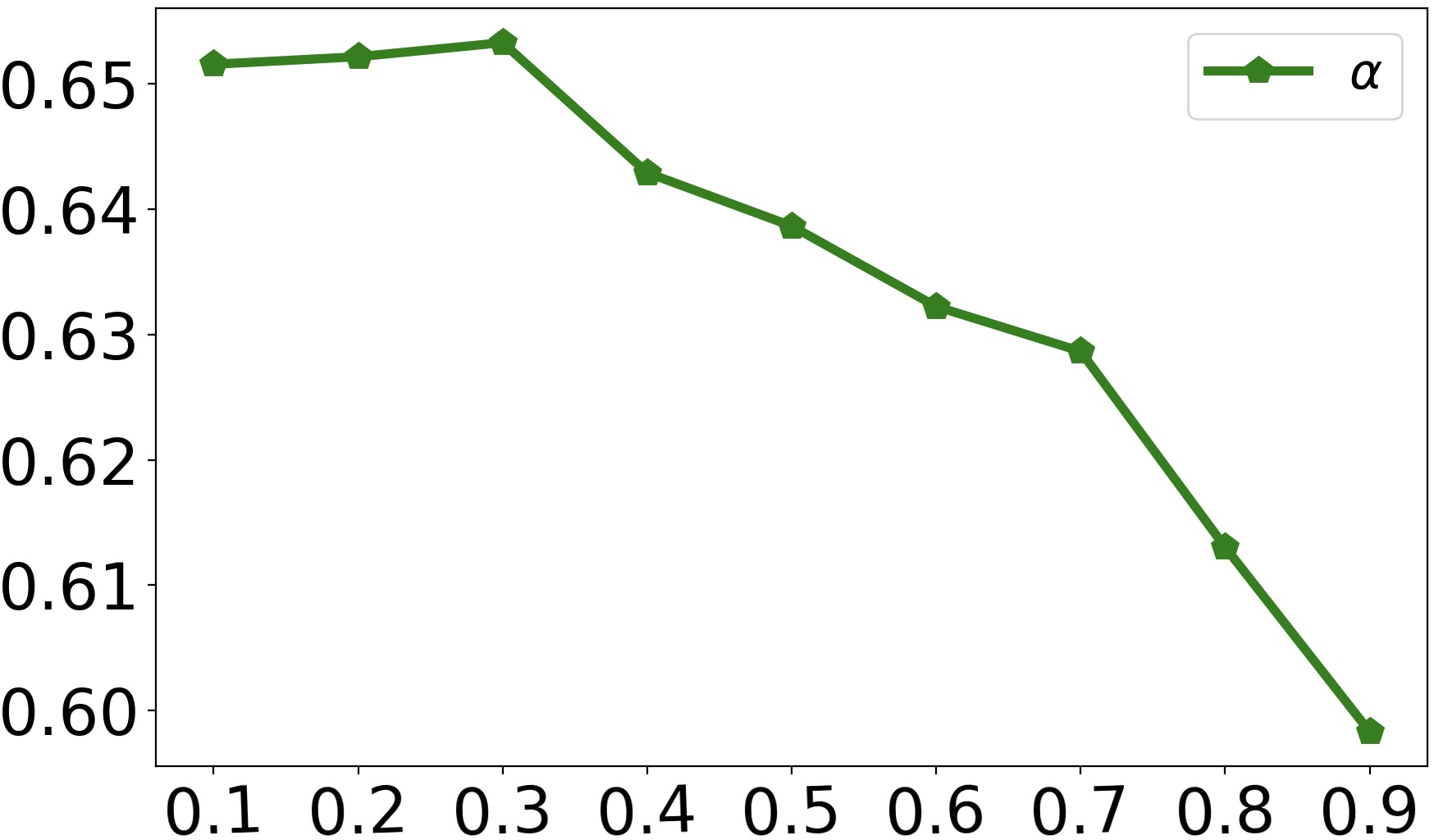}}
\caption{The purchase AUC performance of ablation experiments on the public dataset.}
\label{fig:hyper}
\vspace{-8pt}
\end{figure}

\subsubsection{Hyper-parameter Study (RQ3).}
To explore the sensitivity of CSTWA to perturbations of different hyper-parameters, 
We conduct ablation experiments for hyper-parameters $L, \alpha$, and $\gamma$ on the public dataset.
The results are shown in Figure~\ref{fig:hyper}.

(a) As depicted in Figure~\ref{fig:a}, we vary the values of hyper-parameters $L$ and $\gamma$ to evaluate their impact on the performance. 
We observed that increasing the number of layers for neighborhood propagation do not improve the performance significantly. 
To reduce the computational burden of the graph network, we set $L$ to 1. 
Additionally, we find that setting $\gamma$ to 3 resulted in the best performance, beyond which the performance drop rapidly. 
This indicates that considering instances of information contradiction is crucial, but giving them too much weight is not desirable.

(b) To confirm the efficacy of $\alpha$, we experiment with several different settings, ranging from 0.1 to 0.9, as shown in Figure~\ref{fig:b}.
The best performance is at $\alpha = 3$. 
This suggests that the CTR task’s structure information helps the CVR task. 
However, since CTR and CVR tasks are distinct, setting a higher $\alpha$ leads to poorer results.

\section{Conclusion}
In this work, we elucidate the curse of knowledge problem in CVR methods based on multi-task learning.
For solving the above problem, we propose CSTWA, which contains three effective components (\emph{i.e.},  Structure Migrator, Click Perceptron, and Curse Escaper), which can filter the input information that is more friendly for CVR, calibrate the fundamental representation layer, and reweight the discriminant  layer. Specifically, our method could mine and transfer the task-independent structure information from the auxiliary task (\emph{i.e.}, CTR task).
Meanwhile, it models the click bias information of samples of the entire space.
Besides, it introduces a brand-new weight assignment algorithm to explicitly reinforce CVR-related knowledge.
Extensive experiments on two datasets demonstrate the superior performance of CSTWA.

In the future, we plan to explore more efficient ways of transferring structure information to make our models more cost-effective.

\section{Acknowledgement}
This research was supported by Shopee Live Algorithm team.
\newpage
\section{Ethical Statement}

As authors, we acknowledge the importance of maintaining the integrity of research and its presentation to avoid damaging the trust in the journal, the professionalism of scientific authorship, and ultimately the entire scientific endeavor.   Therefore, we pledge to follow the rules of good scientific practice, which include:

\begin{itemize}
    \item \textbf{Manuscript Submission:} We will not submit the same manuscript to more than one journal simultaneously.
    \item \textbf{Originality:} We will ensure that the submitted work is original and has not been published elsewhere, either partially or in full, in any form or language.   We will provide transparency regarding the reuse of material to avoid concerns about self-plagiarism.
    \item \textbf{Salami Slicing:} We will not split a single study into multiple parts to increase the quantity of submissions and submit them to various journals or to one journal over time.
    \item \textbf{Concurrent Publication:} If we choose to publish concurrently or secondarily, we will meet certain conditions such as translations or manuscripts intended for a different group of readers.
    \item \textbf{Data Presentation:} We will present our results clearly, honestly, and without fabrication, falsification, or inappropriate data manipulation.   We will adhere to discipline-specific rules for acquiring, selecting, and processing data, and we will not present data, text, or theories by others as our own.   Proper acknowledgments will be given for all materials, and we will secure permissions for copyrighted materials.   We understand that the journal may use software to screen for plagiarism.
    \item \textbf{Permissions:} We will obtain permissions for the use of software, questionnaires/(web) surveys, and scales in our studies.
    \item \textbf{Citation:} We will cite appropriate and relevant literature to support our claims in both research and non-research articles, and we will avoid excessive and inappropriate self-citation or coordinated efforts among several authors to collectively self-cite.
    \item \textbf{Truthful Statements:} We will avoid making untrue statements or descriptions about an entity that could potentially be seen as personal attacks or allegations about that person.
    \item \textbf{Public Health and National Security:} We will clearly identify research that may be misapplied to pose a threat to public health or national security.
    \item \textbf{Authorship:} We will ensure that the author group, corresponding author, and order of authors are correct at submission.
\end{itemize}

All of the above guidelines are essential for respecting third-party rights such as copyright and/or moral rights.   As authors, we recognize our responsibility to uphold the highest ethical standards in scientific research and publication.
\newpage

\bibliographystyle{splncs04}
\bibliography{refs}

\end{document}